\documentclass[12pt,twoside]{article}
\usepackage{cmp2e,amssymb,amsbsy,epsf}

\renewcommand{\d}{{\rm d}}

\title[SNRs as cosmic ray accelerators. SNR IC~443]
      {Supernova remnants as cosmic ray accelerators. SNR IC~443}

\author[B.Hnatyk, O.Petruk]{B.Hnatyk, O.Petruk}

\address{Pidstryhach Institute for Applied Problems in Mechanics \& 
	 Mathematics\\
	 National Academy of Sciences of Ukraine\\
	 3$^b$ Naukova St., UA--290601 Lviv, Ukraine}

\date{Received March 15, 1998}

\begin{document}

\maketitle
\setcounter{page}{655}

\begin{abstract}
We examine the hypothesis that some supernova remnants (SNRs) may be 
responsible for some unidentified $\gamma$-ray sources detected by EGRET 
instrument aboard the Compton Gamma Ray Observatory. If this is the case, 
$\gamma$-rays are produced via pion production and decay from direct 
inelastic collisions of accelerated by SNR shock wave ultrarelativistic 
protons with target protons of the interstellar medium. We develop a 3-D 
hydrodynamical model of SNR IC~443 as a possible cosmic $\gamma$-ray source 
2EG~J0618+2234. The derived parameters of IC~443: the explosion energy 
$E_o=2.7\cdot 10^{50}$ erg, the initial hydrogen number density $n(0)=0.21$
cm$^{-3}$, the mean radius $\overline{R}=9.6$ pc and the age $t=4500$ yr 
result in too low $\gamma$-ray flux, mainly because of the low explosion 
energy.  Therefore, we investigate in detail the hydrodynamics of IC~443 
interaction with a nearby massive molecular cloud and show that the reverse 
shock wave considerably increases the cosmic ray density in the interaction 
region.  Meantime, the Rayleigh-Taylor instability of contact discontinuity 
between the SNR and the cloud provides an effective mixing of the containing 
cosmic ray plasma and the cloud material. We show that the resulting 
$\gamma$-ray flux is consistent with the observational data.

\pacs 	98.38.Mz, 	
	98.70.Qy, 	
	98.70.Rz, 	
	98.70.Sa	
\end{abstract}

\section{Introduction}

Supernova remnants (SNRs) are believed to be the most promising accelerators 
of ultrarelativistic particles (electrons, protons and nuclei) in our Galaxy 
\cite{Berez-Bulanov90,Gaisser90}.  
They are responsible for the majority of cosmic rays (CRs) with energies up to 
$10^{15}\ eV$ (Lorentz factor of proton $\gamma_p \sim 10^6$).  Classical 
observations of synchrotron radiation from SNRs, mainly in the radio band, give 
experimental confirmation of the presence of relativistic electrons in 
these cosmic objects \cite{Lozins}. Recently, the presence of ultrarelativistic electrons in 
SNRs has been confirmed by optical, X-ray and soft $\gamma$-ray observations 
\cite{Lozins,Reynolds96}. \par

There are expected two main CR accelerators in SNR. One of them is a 
young pulsar (if there is any), i.e., a spinning neutron star with a strong 
magnetic field. The second one is a strong shock wave as the result of an 
interaction of supernova ejecta with the interstellar medium (ISM). It is the 
main object of our investigation.  \par

The theory of CR acceleration at the shock front via the first order Fermi 
mechanism is quite well elaborated 
\cite{Berez-Bulanov90,Gaisser90,Heavens84}. 
Its main prediction is a 
considerably higher efficiency of the proton acceleration in comparison with 
an electron case, i.e., up to $10\%$ of kinetic energy of the incoming flow may be 
transformed into the energy of newly accelerated CRs. Experimental 
evidences of the proton acceleration at shock fronts are limited to the 
solar system region (a bow shock at the boundary of the Earth's 
magnetosphere, interplanetary shocks 
and solar flares). Therefore, it is of great importance to find additional 
evidences of the proton and nucleus component acceleration at the fronts of cosmic 
shock waves. \par

Such a new possibility is offered by the recent observations of considerable 
$\gamma$-ray fluxes in directions on some radio bright SNRs 
\cite{Sturner-Dermer95,Esposito-Hunter96,Sturner-Dermer-M96}. 
Among the 32 
unidentified Galactic plane $\gamma$-ray sources detected by the Energetic 
Gamma-Ray Experimental Telescope (EGRET) aboard the Compton Gamma-Ray 
Observatory (CGRO) five sources are coincident with SNRs without known 
pulsars. Four of these SNRs (W28, W44, $\gamma$ Cygni and IC~443) exhibit 
evidences of a strong interaction with interstellar clouds. Therefore, it is 
believed that the observed $\gamma$-ray emission is the result of an interaction 
of the accelerated by SNR shock wave protons with the SNR matter itself 
\cite{Sturner-Dermer95} or with the matter of the adjacent molecular clouds 
\cite{Esposito-Hunter96}. \par

In this paper we check this hypothesis for the case of well-studied SNR IC~443 
and show that according to the atypically low initial energy of a supernova 
explosion such simple explanations do not hold here. In order to explain the 
$\gamma$-ray emission of IC~443 we elaborate a hydrodynamical model of 
IC~443 as a result of a supernova explosion in the medium with a large scale 
density 
gradient, what allows us to explain observational data concerning X-ray 
surface brightness distribution and obtain the parameters of remnants, 
especially shock wave characteristics (section \ref{sect2_paperV}). 
Using these parameters, we consider in 
detail the SNR shock interaction with a nearby molecular cloud 
(section \ref{sect3_paperV})
and show that the 
reverse shock enhances the energy and number density of newly accelerated 
CRs (section \ref{sect4_paperV}). 
Then we show that caused by the Rayleigh-Taylor instability, 
effective mixing of SNR and molecular cloud matter leads to the generation of 
an observable flux of $\gamma$-rays (section \ref{sect5_paperV}). Conclusions 
are given in section \ref{sect6_paperV}.\par

\section{Hydrodynamical model of SNR IC~443}
\label{sect2_paperV}

\begin{figure}[t]
\unitlength=1pt
\begin{centering}
\begin{picture}(397,397)
\epsfxsize=140mm
\put(0,0){\epsffile[212 336 384 508]{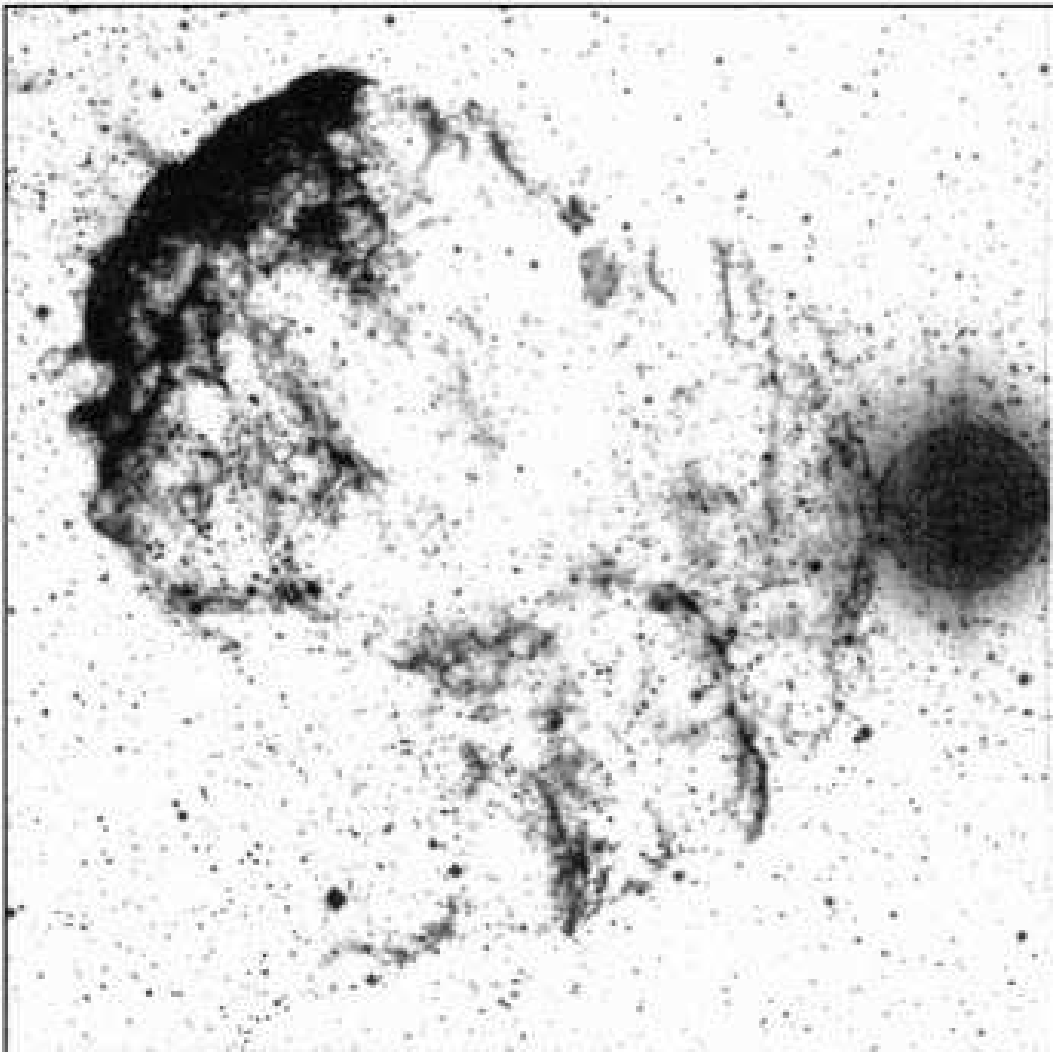}}
\end{picture}\\
\end{centering}
\caption{Palomar Sky Survey red plate image of IC~443.}
\label{ic433_red}
\end{figure}

\begin{figure}[t]
\unitlength=1pt
\begin{centering}
\begin{picture}(310,310)
\epsfxsize=110mm
\put(0,0){\epsffile[210 336 384 508]{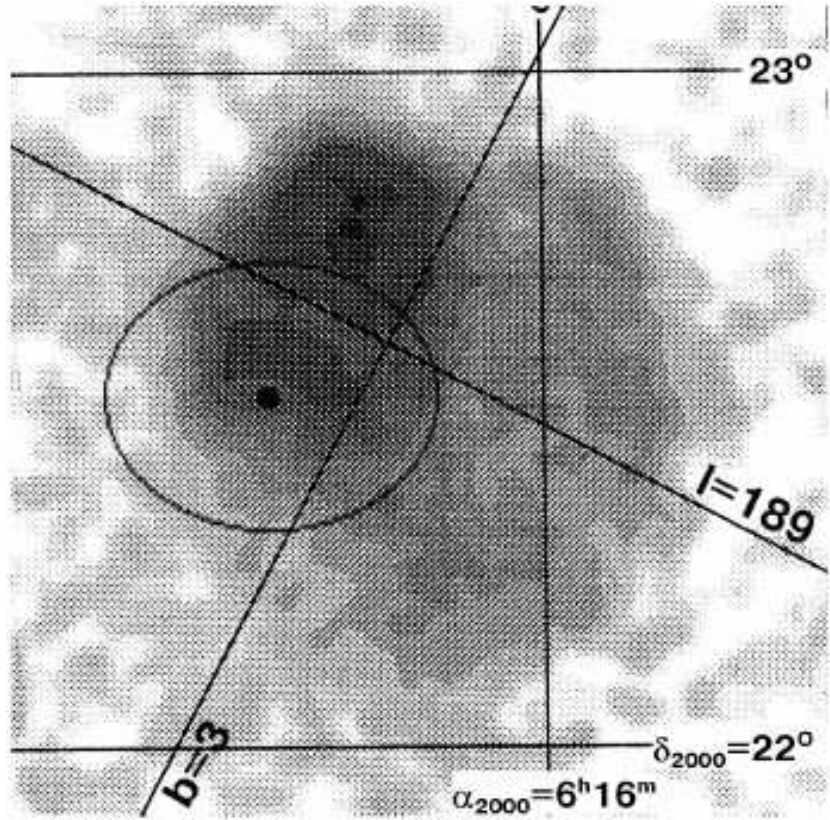}}
\end{picture}\\
\end{centering}
\caption{X-ray image of IC~443 obtained during the ROSAT all-sky survey 
\protect\cite{Asaoka-Asch94} and 95\% confidence contour (solid line) of 
$\gamma$-ray source 2EG~J0618+2234 \protect\cite{Esposito-Hunter96}.}
\label{ic433_xrays}
\end{figure}

SNR IC~443 is well studied at different energy bands from radio to 
$\gamma$-rays 
\cite{Green86,Braun-Strom86,Mufson-McCl86,White-Rainey87,Petre-SzymkS88,Asaoka-Asch94} 
(figure~\ref{ic433_red} and \ref{ic433_xrays}). 
It is believed to be the result of a supernova explosion 
which happened about $t=3000\div 5000$ yr ago near a dense molecular 
cloud at the distance about $d\approx 1.5$ kpc from us. Its unusual nonspherical 
morphology should reflect the inhomogeneity of the surrounding ISM, but until 
now a self-consistent model of IC~443 is absent. 
We suppose that the main reason of this is the absence of a detailed 
numerical modelling of IC~443. So, despite the evident influence of 
inhomogeneity of the surrounding ISM,  IC~443 models are mainly based on the 
use of the one-dimensional (1D) self-similar Sedov solution \cite{Sedov} for 
SNR evolved in a uniform medium. The role of the density gradient in the 
surrounding ISM was taken into account only for the explanation of shape 
asymmetry within the framework of the approximate Kompaneetz \cite{Komp} 
approach \cite{Lozins,Petre-SzymkS88}. 

Therefore, we carry out here a detailed 2D calculation of the IC~443 
evolution in a nonuniform medium using the proposed in 
\cite{Hn88,Hn-Pet-96,Hn-Pet98} new approximate method of a complete 
hydrodynamical description of a point-like explosion in an arbitrary 
nonuniform medium. Besides geometrical sizes, we took the total flux and the 
surface brightness distribution in an X-ray band as free characteristics 
which give the best possibility for the estimation of plasma characteristics 
inside the SNR.
\clearpage

\begin{figure}[t]
\unitlength=1pt
\begin{centering}
\begin{picture}(397,177)
\epsfxsize=140mm
\put(0,0){\epsffile[89 330 504 515]{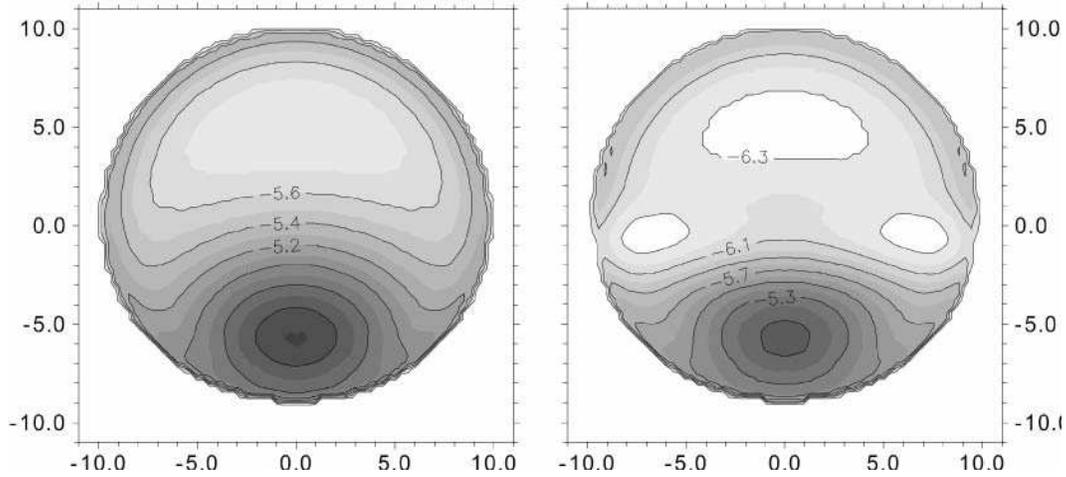}}
\end{picture}\\
\end{centering}
\caption{Surface brightness $S_x$ distribution for the model of IC~443 in the
ROSAT X-ray ($0.1\div 2.4$ keV) band. The angle between the symmetry axis of 
the SNR and the line of sight is $45^o.$ The values of $S_x$ in erg $s^{-1}$ 
cm$^{-2}$st$^{-1}$ are given in the figure. (a) Model with no absorption 
correction. (b) Model which includes interstellar and molecular cloud 
absorption.} 
\label{ic433_xrays_models}
\end{figure}

Taking into account the generally accepted distance to IC~443 $d=1.5$ kpc 
\cite{Lozins,Petre-SzymkS88,Asaoka-Asch94}, 
we have reproduced the following picture of the supernova explosion. 
At $t=4500$ yr ago, a SN explosion with the total energy $E_o=2.7\cdot 
10^{50}$ erg took  place in the transition zone between a warm ISM with the 
hydrogen number density $n_{o}$ and an interstellar HI cloud with the number 
density $n_{i}$ and the high scale $h,$ so the initial density distribution 
in this transition zone was 
$
\displaystyle
n(\tilde r)=n_{o}+n_{i}\exp(-\tilde r/h),  
$
with $n_{o}=0.16$ cm$^{-3}$, $n_{i}=16$ cm$^{-3}$, $h=2.4$ pc, the 
SN position moved on $\tilde r=r_o=13.8$ pc 
from the cloud centre and the initial hydrogen 
number density in the point of explosion $n(0)=0.21$ cm$^{-3}.$ 
Calculations show the following best fitting parameters of the IC~443 model.  
Depending on the direction from the explosion centre, the shock radius 
$R,$ its velocity $D,$ the postshock temperature $T$ and the initial 
hydrogen number density at shock front position vary in the 
ranges: $7.5\leqslant R$ (pc) $\leqslant 10.1,$ 
$480\leqslant D$ (km/s) $\leqslant 880,$ $3.1\cdot10^6\leqslant T$ (K) $\leqslant 
1.1\cdot 10^7,$ 
$0.16\leqslant n$(R) (cm$^{-3})\leqslant 1.31$. The total swept up mass of the 
ISM is  
$M_{SNR}=28.0$ masses of the Sun ($M_\odot$), the volume of the disturbed by a 
shock 
wave region is $V=1.1\cdot10^{59}$ cm$^3$, the mean SNR radius is 
$\overline{R}=(3V/4\pi)^{1/3}=9.6$ pc. 
In figure~\ref{ic433_xrays_models} the X-ray surface brightness of the IC~443 
model is presented. 
The total equilibrium flux in an X-ray band 
is calculated for X-ray emissivity data from \cite{Raym-Smith77} and is 
obtained to be $L_x(>2.4$ keV)=1.4$\cdot 10^{34}$ erg/s  
(no absorption correction is made) which corresponds to the experimental result 
$L_{x, obs}(>2$ keV)=1.5$\cdot 10^{34}$ erg/s \cite{Petre-SzymkS88}. 
In figure~\ref{ic433_xrays_models}(b) we show 
the effect of the absorption of interstellar gas with the column density 
$N_{H,\ ISM}=3\cdot 10^{21}$ cm$^{-2}$ and an 
elongated cylindrical-like molecular cloud in front of the SNR 
with the radius $R_{cl}=2$ pc and the cloud column density 
$N_{H,\ cl}=(0\div5)\cdot 10^{21}$ cm$^{-2}.$ 
The data for the absorption cross-section from \cite{Morr-McCammon83} 
were used. This absorption is 
mainly responsible for the unusual nonspherical X-ray shape of IC~443.

\section{Hydrodynamics of the SNR IC~443 -- molecular cloud interaction}
\label{sect3_paperV}

Our calculations show that the visible X-ray shape of IC~443 is close to spherical,  
although with different radii in the south-west and the north-east, while the 
ROSAT X-ray image provides a strong evidence of two half-spheres in the IC~443 
structure \cite{Asaoka-Asch94}. 
This difference may be naturally explained by the enhanced absorption 
of X-rays in the adjacent to IC~443 elongated molecular cloud which is visible 
in a radio band \cite{Braun-Strom86,Mufson-McCl86,White-Rainey87,Dickman-Snell92}.  
\par

The same molecular cloud (or, more correctly, the part of the cloud 
disturbed by the shock wave of SNR) should be responsible for the observable 
$\gamma$-rays. Observations in ${\rm CO}$ and ${\rm HCO^+}$ emission 
radiolines \cite{Dickman-Snell92} 
reveal that the total mass disturbed by the shock gas in the molecular cloud 
reaches 
$500\div 2000\ M_\odot.$ Although this molecular cloud covers practically the 
entire remnant, the interaction region is localized in the southeastern part and 
coincides with the $\gamma$-ray source position (figure \ref{ic433_xrays}). 
It means that the interaction region is somewhat displaced from the SNR symmetry 
axis (on the angle $0\div 30^o$, as viewed from the Supernova progenitor 
position), i.e., IC~443 has essentially a 3D shape. \par

From the hydrodynamical point of view, the SNR -- molecular cloud 
interaction is a complicated event accompanied by the arising of two shocks: 
a forward shock in the cloud matter and a reverse shock in the SNR matter. 
The boundary between the ISM and the 
molecular cloud becomes a contact discontinuity between the shocked plasmas of 
the SNR and the cloud \cite{Lozins}. 
According to the complexity of the resulting flow we cannot 
calculate it within the framework of the used above method for one shock flow 
calculations. But simple analytical estimates are 
possible. So, the velocity of the forward shock wave in the molecular cloud $D_m$ 
is determined by the SNR shock velocity $D_i$ at the cloud position and the 
ratio of hydrogen number densities of the molecular cloud $n_m$ and the ISM $n_i$ 
\cite{Lozins}: 
$                     
\displaystyle
D_m=D_i\sqrt{\beta n_i/n_m},
$
where $\beta$ depends on the number density contrast and $\beta=6$ for 
$n_m/n_i\to \infty.$ If a SNR shock wave penetrates into the molecular cloud at 
the time $t_p\sim 3500$ yr after the SN explosion,   
when the shock radius is $R_p\sim 7$ pc, $D_i\sim 600$ km/s 
and, according to the Sedov solution 
\cite{Sedov}, the effective thickness of the postshock gas shell is 
$\Delta R_p\sim 0.1R_p,$ the typical time scale of the reverse shock wave existence 
is 
$$
\Delta t_{\rm rev}\sim{\Delta R_p\over D_i}=1.2\cdot 10^3\left({R_p\over 
7\ {\rm pc}}\right)\left({D_i\over 600\ {\rm km/s}}\right)^{-1}\quad {\rm yr}.
$$
The forward shock penetrates into the molecular cloud up to the distance 
$
\displaystyle
L_p=D_m\cdot\Delta t_{\rm rev}\sim 0.1 R_p\sqrt{\beta n_i/n_m},
$
and the shock-excited mass of the cloud is of order 
$$
M_m=L_p\cdot\rho_m\cdot S_m\sim 7.2\cdot 10^2
\left({R_m\over 5\ {\rm pc}}\right)^2
\left({R_p\over 7\ pc}\right) 
\left({n_i\over 0.3\ cm^{-3}}\right)^{1/2} 
\left({n_m\over 10^5\ cm^{-3}}\right)^{1/2} 
\quad M_\odot,
$$
where $\rho_m=1.4n_mm_p$ is the molecular cloud density (the typical 
number density of helium atoms $n_{He}=0.1n$ is taken), 
$m_p$ is the proton mass, $S_m=\pi R^2_m$ is the 
effective surface of the SNR -- cloud interaction.
We note that the obtained mass of the shocked matter of the molecular cloud is 
close to the observed one, if we take typical values $n_m=10^4\div 10^5$ 
cm$^{-3}$.\par

To summarise, the proposed here 3D model of IC~443 explains in a consistent way 
the main characteristics of the remnant, including its X-ray properties and 
interaction with the adjacent molecular cloud. We use this model for the 
estimation of the efficiency of cosmic ray acceleration and for the explanation 
of $\gamma$-ray production in IC~443.

\section{Acceleration of cosmic rays in SNR IC~443}
\label{sect4_paperV}

CRs, i.e., high energy relativistic particles (electrons, 
protons, nuclei) with the power-law energy spectrum $N\propto \varepsilon^{-\alpha},\ 
\alpha\sim2.6\div3.2$ and the mean energy density $\omega_{cr}\sim 
1$ eV/cm$^3\sim10^{-12}$ erg/cm$^3$ are believed to be accelerated by 
different mechanisms in our Galaxy (CRs with energies $\varepsilon\leqslant 
10^{17}$ eV) and in powerfull extragalactic sources (CRs with energies 
$\varepsilon\geqslant 10^{17}$ 
eV, up to the maximal observable energies $\varepsilon_{\max}=3\cdot10^{20}$ 
eV), such as active galactic nuclei, quasars, radiogalaxies etc. By now the most 
promising mechanism for the galactic CR generation is the first order Fermi 
mechanism at shock wave fronts 
\cite{Berez-Bulanov90,Gaisser90,Heavens84}. 
In this mechanism fast particles 
gain energy during their diffusive motion in the shock region 
as a result of scattering on different types of magnetic field 
fluctuations (magnetohydrodynamical waves, especially $\rm Alfv\acute en$ 
waves, turbulent pulsation etc.) ahead and behind the shock front. 
The main reason for 
effective energy gain is the convergent character of hydrodynamical flows 
in the shock front frame. \par

The main characteristics of a shock wave acceleration are the following
\cite{Berez-Bulanov90,Gaisser90}. The acceleration time is
{\small
$$
t_{\rm acc}=
{3\over u_1-u_2}\left({k_1\over u_1}+{k_2\over u_2}\right)=
2.1\cdot10^3\left({\eta\over10}\right)\left({D\over 10^3 {\rm km/s}}\right)^{-2}
\left({H_1\over 10^{-5}\ G}\right)^{-1}\left({\varepsilon\over 
10^{12}\ {\rm eV}}\right) \ {\rm yr},
$$
} 
where $u_i$ ($i=1$ for upstream and 2 for downstream) is the plasma velocity
in the shock frame, $k_i=l_ic/3$ is the diffusion coefficient, 
$l_i=\eta r_{L,i}$ is the mean free path, $r_{L,i}=\varepsilon/eH_i$ 
is the Larmor 
radius of a particle with the energy $\varepsilon$ and the charge $e$ 
in the magnetic field $H_i,$ 
$c$ is the light velocity, $D=-u_1$ is the shock velocity in the observer's frame. 
Parameter $\eta$ is believed to be in the range $1\leqslant\eta\leqslant 10$ 
for SNR shocks. 

The theoretically predicted spectrum of accelerated particles has the 
power-law type
\begin{equation}
N(\varepsilon)\d\varepsilon=N(\varepsilon_o)\cdot 
(\varepsilon/\varepsilon_o)^{-\alpha}\d\varepsilon\qquad 
{\rm with}\quad \alpha=3u_1/(u_1-u_2)-2,
\label{spectum_of_CR}
\end{equation}
so $\alpha=2$ for plasma with the adiabatic index $\gamma_{\rm ad}=5/3.$\par

Both electrons and protons are involved in the acceleration process but the 
number of accelerated electrons is considerably less than the number of protons
$
\displaystyle
{N_e(\varepsilon)/N_p(\varepsilon)}=\left(m_e/m_p\right)^{(\alpha-1)/2}.
$
Further we shall consider only a proton component of SNR CRs.\par

There are different estimates of the efficiency of energy transformation from  
the kinetic energy of a hydrodynamical flow into the energy of accelerated 
particles \cite{Berez-Bulanov90,Gaisser90,Heavens84}. 
The reasonable result is that about 10\% of the 
kinetic energy of a flow may be transformed at the shock front into the 
energy of CRs.The maximal energy of CRs is restricted by the processes of 
energy loss, particle escape from the region of acceleration, finite time of 
the shock existence etc. In the considered here SNR IC~443 case, taking into 
account the calculated IC~443 shock characteristics, we should expect the 
following characteristics of CR acceleration. The total energy of cosmic 
rays in IC~443 is 
$$
W_{\rm cr}=\nu E_o=10^{48}\left({\nu\over0.01}\right)\left({E_o\over10^{50}\ 
{\rm erg}}\right)\quad {\rm erg}
$$
(in order to provide the necessary amount of CRs in our Galaxy due to SNRs we 
need $\overline{\nu}=0.03$). So, for our case $E_o=2.7\cdot 10^{50}$ erg we 
obtain $W_{cr}=8.1\cdot10^{48}.$\par

For the spectral index of protons we take $\alpha=2.1$ which is close to 
theoretical predictions (\ref{spectum_of_CR}) 
and corresponds to the observable 
$\gamma$-ray spectrum (see below). Then the energy spectrum of IC~443 CRs 
(mainly protons) is 
{\footnotesize
\begin{equation}
N_P(\varepsilon)=
{(\alpha-2)W_{\rm cr}\over \varepsilon_o^2}\left({\varepsilon\over 
\varepsilon_o}\right)^{-\alpha}=
2.3\cdot 10^{47}\left({\nu\over 0.01}\right)
\left({E_o\over 10^{50}\ {\rm erg}}\right)
\left({\varepsilon\over 600\ {\rm MeV}}\right)^{-2.1}\quad {{\rm prot}\over 
{\rm MeV}}.
\label{en_spectrum_CR_IC443}
\end{equation}
}

The process of the pion creation in subrelativistic proton -- proton at rest 
collisions is effective beginning from the proton kinetic energy 
$\varepsilon=\varepsilon_o\approx 600$ MeV for which the pion decay 
dominates over other mechanisms of $\gamma$-ray production 
\cite{Sturner-Dermer-M96}.
The same proton energy is needed to generate $\gamma$-photons with the 
energy $\varepsilon_\gamma\geqslant 100$ MeV. 
The total number 
of such protons in IC~443 is 
$$
N_{\rm tot}(>600\ MeV)={\alpha-2 \over \alpha-1}{W_{cr}\over 
\varepsilon_o}=0.93\cdot
10^{50}\left({\nu\over 0.01}\right)
\left({E_o\over 10^{50}\ {\rm erg}}\right)\quad {\rm prot}.
$$
The maximal energy of CRs is restricted in the IC~443 case by the SNR age 
$t=4500$ yr. Then, from $t=t_{\rm acc}$ we obtain 
{\small
$$
\varepsilon_{\max}=
{3\over 20}{eH\over\eta}\left({D\over c}\right)^2t=
0.48\left({t\over 10^3\ {\rm yr}}\right)\left({\eta\over 10}\right)^{-1}
\left({D\over 10^3\ {\rm km/s}}\right)^2\left({H\over 10^{-5}\ G}\right)\quad 
{\rm TeV}.
$$
}

The mean energy density of CRs inside IC~443 with volume $V$ is
\begin{equation}
\overline{\omega}_{\rm cr}(\overline{R})=
{W_{\rm cr}\over V}=5.2\left({\nu\over 0.01}\right)
\left({E_o\over 10^{50}\ {\rm erg}}\right)\left({\overline{R}\over 
10\ {\rm pc}}\right)^{-3}
\quad {{\rm eV}\over {\rm cm}^{3}}.
\label{mean_energy_dens_CR_IC443}
\end{equation}
Exactly speaking, CRs should not be uniformly distributed inside IC~443. 
Rather they will be concentrated in a thin shell of thickness $\Delta R\sim 
0.1R$ $(\Delta V\simeq 0.3{\rm V})$ near the shock front, where the majority of 
SNR mass and about half of its thermal energy are concentrated 
\cite{Sedov}, since the 
necessary time for diffusion on the distance of the order of a shock 
radius
$$
t_{\rm d}={\overline{R}^2\over k_2}=10^7\eta^{-1}\left({\overline{R}\over 
10\ {\rm pc}}\right)^2
\left({\varepsilon\over 10^{12}\ {\rm eV}}\right)^{-1}\left({H\over 
10^{-5}\ G}\right)
\quad {\rm yr}
$$
considerably exceeds the IC~443 age. It means that at least half of the total 
energy of CRs $W_{\rm cr}$ is concentrated in this shell, so, 
the CR energy density near the shock front is 
$
\displaystyle
\omega_{\rm sh}\approx0.5W_{\rm cr}/0.3{\rm V}\approx 1.7\overline{\omega}_{\rm cr}.
$
As we shall see below, this value is too low for the generation of the necessary 
gamma-flux from IC~443. Therefore, we propose and 
consider in detail the promising 
mechanism of the CR energy density enhancement in IC~443, connected with the 
reverse shock action.\par

In the reverse shock the plasma of SNR is subjected to another shock and 
its final values of the number density $n_f=\gamma_{\rm ad}/(\gamma_{\rm ad}-1)n_2$ and 
magnetic field $H_f=(n_f/n_2)H_2$ are 2.5 times enhanced for 
$\gamma_{\rm ad}=5/3$ plasma \cite{Landau-Lifshitz}, 
which results in the increase of CR number density 
$n_{{\rm cr},f}/n_{{\rm cr},2}=n_f/n_2=2.5.$ Meantime the energy of individual 
ultrarelativistic particles increases according to adiabatic invariant  
conservation ($P^2_\bot /H=\mbox{const}$ where $P_\bot=m_pv_\bot$ is the 
normal to the magnetic field orientation component of particle's momentum) 
$\varepsilon_f/\varepsilon_2=\sqrt{H_f/H_2}$. Therefore, we expect that behind 
the reverse shock front the energy density of CRs will be 
{\small
\begin{equation}
\omega_{\rm rev}=
\omega_{\rm sh}\ \left({n_f\over n_2}\right)^{3/2}= 
6.6\ \overline{\omega}_{\rm cr}(R_p)=
101\left({\nu\over 0.01}\right)
\left({E_o\over 10^{50}\ {\rm erg}}\right)\left({R_p\over 7\ {\rm pc}}\right)^{-3}
\quad {{\rm eV}\over {\rm cm}^{3}}.
\label{energy_dens_CR_IC443}
\end{equation}
}
For our model we will have $\omega_{\rm rev}=818$ eV/cm$^{3}$. 
The maximal energy of CRs in the reverse shock 
region will be of the order of maximal energy in the main shock, i.e. about 
$10^{12}$ eV, while acceleration conditions in the both shocks are similar.

\section{{$\boldsymbol\gamma$}-ray emission from IC~443}
\label{sect5_paperV}

The $\gamma$-ray flux from source 2EG~J0618+2234 which coincides with the 
SNR IC~443 position is well approximated by the power-law dependence 
\cite{Merck-Bertsch96} 
$$
S_\gamma=(5.4\pm 0.4)\cdot10^{-10}\left({\varepsilon_\gamma\over 293\ 
{\rm MeV}}\right)^{-2.1\pm0.1}\quad {{\rm phot}\over {\rm cm}^2\cdot {\rm s}
\cdot {\rm MeV}}
$$
in the EGRET band $20\ {\rm MeV}\leqslant \varepsilon_{\gamma}\leqslant 
30$ GeV.
For the distance to IC~443 $d=1.5$ kpc the $\gamma$-ray spectrum of the whole 
SNR is 
$$
F_\gamma=4\pi d^2S_\gamma=1.4\cdot 10^{36}\left({\varepsilon_\gamma\over 100\ 
{\rm MeV}}\right)^{-2.1}\quad {{\rm phot}\over {\rm s}\cdot {\rm MeV}}.
$$
The total $\gamma$-ray luminosity in $\varepsilon_\gamma\geqslant100\ {\rm MeV}$ 
band is $L_\gamma,obs=2.2\cdot 10^{35}$ erg/s and the rate of photon creation is 
$\dot{N}_{\gamma ,{\rm tot}}=7.6\cdot 10^{38}$ phot/s.\par

In \cite{Sturner-Dermer95} it was shown that in the case of a pulsar 
absence, the most promising mechanism for $\varepsilon_\gamma>100\ {\rm 
MeV}$ $\gamma$-ray production is inelastic interaction of relativistic 
protons with protons at rest resulting in the creation of pions and their 
consequent decay into $\gamma$-rays. The $\gamma$-ray luminosity in this case 
is equal to the rate of energy transformation from relativistic protons to 
neutral pions \cite{Berez-Bulanov90}
\begin{equation}
L_\gamma=c\ n_N\int\limits_{\varepsilon_{\min}}^\infty 
N(\varepsilon)\sigma_{pp}(\varepsilon)
\overline{\varepsilon}_{\pi^o}(\varepsilon)\d\varepsilon
\label{gamma_lum_IC443}
\end{equation}
where $n_N=1.4n$ is the mean number density of target nuclei in the region of 
interaction, $\varepsilon_{\min}\approx600$ MeV 
is the minimal proton kinetic energy 
of the effective pion creation (with the cross-section 
$\sigma_{pp}(\varepsilon)$  
close to the mean value $\overline{\sigma}_{pp}=3\cdot10^{-26}$ cm$^2$), 
$\overline{\varepsilon}_{\pi^o}(\varepsilon)=\varepsilon/6$ 
is the mean energy transformed into the pion.\par

Substituting in (\ref{gamma_lum_IC443}) the cosmic ray 
spectrum (\ref{en_spectrum_CR_IC443}) we obtain: 
\begin{equation}
L_\gamma={c\overline{\sigma}_{pp}\over 6}n_N W_{\rm cr}
=0.2\cdot 10^{35}n\left({W_{cr}\over 10^{50}\ {\rm erg}}\right) 
\quad {{\rm erg}\over {\rm s}}~.
\label{gamma_lum_theor}
\end{equation}
So, as it was mentioned earlier in \cite{Sturner-Dermer95}, it is necessary 
to have the initial number density in the IC~443 region $n\sim 10$ cm$^{-3}$ and 
the total energy of CRs of $W_{\rm cr}\sim 10^{50}$ ergs in order to explain 
the observable $\gamma$-ray flux in such a simple model. But, as we showed 
in section \ref{sect2_paperV} in the IC~443 case 
supernova exploded in a low density 
region with $n\sim 0.3$ cm$^{-3}$ and the total energy of CRs is expected to 
be only of the order of $W_{\rm cr}\sim 10^{48}$ erg. \par 

In \cite{Esposito-Hunter96}, the $\gamma$-flux 
from IC~443 is explained as the result of the interaction of accelerated by SNR CRs 
with the matter of the adjacent molecular cloud. The role of the 
molecular cloud consists in the increase of the number density of 
target nuclei $n_N$ 
in the interaction region. Namely, we can rewrite (\ref{gamma_lum_theor}) 
in the form 
\begin{equation}
L_\gamma={c\overline{\sigma}_{pp}\over 6}n_N W_{\rm cr} 
\approx{c\overline{\sigma}_{pp}\over 6}{\omega_{\rm cr}M_{\rm gas}\over m_{\rm p}}
\quad {{\rm erg}\over {\rm s}}
\label{gamma_lum_theor2}
\end{equation}
where $M_{\rm gas}$ is the total mass of the interacting with CRs gas, 
$m_{\rm p}$ is the proton mass.\par

In \cite{Esposito-Hunter96}, 
this total mass is taken to be $M_{\rm gas}=5\cdot 10^{3}\ M_\odot$ 
and the deduced mean CR energy density inside IC~443 is 
$\overline{\omega}_{\rm cr}=96$ eV/cm$^{3}$. 
But, as we can see from (\ref{mean_energy_dens_CR_IC443}), this value of 
$\overline{\omega}_{\rm cr}$ is considerably higher than the one expected for 
IC~443. Another problem here is a high total mass of clouds $M_{\rm gas}\sim 
5\cdot 10^{3}\ M_\odot$ in the region of interaction, while the cloud mass 
outside SNR is included (as we mentioned above, only $500\div 2000 M_\odot$ 
of gas is inside the SNR). The energy density of CRs outside the SNR should 
be considerably  lower than that inside it 
\cite{Berez-Bulanov90,Heavens84}, and, therefore, the pion 
production should be depressed. \par

We suppose that the observable $\gamma$-ray flux  from IC~443 is caused by 
enhancement of the CR energy density in the reverse shock from the SNR -- 
molecular cloud interaction. Indeed, using (\ref{energy_dens_CR_IC443}) 
we obtain from (\ref{gamma_lum_theor2}) the 
following $\gamma$-ray luminosity in the case of enhancement of the CR 
energy density:
\begin{equation} 
\begin{array}{l}
L_\gamma=
{\displaystyle
6.8\cdot10^{-2}{\sigma_{pp}\ c\over m_{\rm p}}
\ \left({n_f\over n_2}\right)^{3/2}M_{\rm gas}W_{\rm cr}R_{\rm p}^{-3}
}\\ \\
{\displaystyle
\quad\  =2.9\cdot10^{34}
\left({W_{\rm cr}\over10^{48}\ {\rm erg}}\right)
\left({M_{m}\over10^3\ M_\odot}\right)
\left({R_{\rm p}\over7\ {\rm pc}}\right)^{-3}
\quad {{\rm erg}\over {\rm s}}.
}
\label{new_L_gamma}
\end{array}
\end{equation}
Therefore, we obtain the theoretical $\gamma$-ray luminosity 
$L_{\gamma}=2.3\cdot10^{35}$ erg/s for IC~443 
despite the low explosion energy $E_o=2.7\cdot 10^{50}$ erg and, 
accordingly, the low total energy of cosmic rays 
$W_{\rm cr}=\overline{\nu}E_o=8.1\cdot 10^{48}$ erg. \par

Additional support for this model follows from the existence of an effective 
mixing mechanism of a hot containing CR plasma and dense shocked plasma of 
the molecular cloud -- the well known Rayleigh-Taylor instability of 
a contact discontinuity between the both media which should lead to a decay 
of contact discontinuity and fragmentation of the shocked cloud material 
onto a set of cloudlets \cite{White-Rainey87}. \par

Another possibility of the effective CR -- cloud material interaction takes place 
in the case of a strongly nonuniform cloud, in which dense molecular cores of mass 
$\sim 10\ M_\odot$ and number density $\sim 10^5$ cm$^{-3}$ are embedded 
in a relatively less dense intercloud medium with the number density 
$\sim 10\div10^2$ cm$^{-3}.$ 
As a result of displacement of a contact discontinuity in the molecular 
cloud direction, such separate cores penetrate into the shocked by a reverse 
shock plasma region and should be the sites of an intense $\gamma$-ray 
production. Radio observations of the shocked molecular gas 
in the IC~443 region reveal the existence of few such cloudlets 
\cite{Dickman-Snell92}.

\section{Conclusions}
\label{sect6_paperV}

If the association of some unidentified EGRET sources with SNRs is 
confirmed, it will be a new argument for the CR acceleration in SNRs. It is 
of great importance that contrary to a radio band, $\gamma$-rays give 
unique information about the acceleration of a proton component of CRs in SNRs. 
In this work we have proposed a self-consistent model of IC~443 which 
explains a total set of its main parameters, including shape, size, 
X-ray and $\gamma$-ray fluxes. We have shown that, despite the low 
supernova explosion energy, account for the SNR -- molecular cloud 
interaction, especially for the role of the reverse shock wave, is the main 
reason of an intense $\gamma$-ray flux from IC~443. Further observational 
and theoretical efforts are needed for checking the theory of CRs 
acceleration by SNRs and IC~443 is a prominent candidate for such 
investigations. \par


\label{last@page}
\end{document}